\documentclass[preprint,12pt]{elsarticle}
\usepackage{amssymb}
\usepackage{soul}
\usepackage{amsmath}
\usepackage{graphicx}
\usepackage{natbib}
\usepackage{color}

\usepackage[figuresright]{rotating}

\journal{Physics Letters A}

\begin{document}

\begin{frontmatter}




\title{Approximate rogue wave solutions of the forced and damped Nonlinear 
Schr\"odinger equation for water waves}


\author{Miguel Onorato and Davide Proment}

\address{Dipartimento di Fisica, Universit{\`a} degli Studi di Torino, Via Pietro Giuria 1, 10125 Torino, Italy, EU}
\address{INFN, Sezione di Torino, Via Pietro Giuria 1, 10125 Torino, Italy, EU}

\begin{abstract}
We consider the effect of the wind and the dissipation on the nonlinear stages of the modulational 
instability. By applying a suitable transformation, we map  the forced/damped Nonlinear Schr\"{o}dinger (NLS)
equation  into the standard NLS with constant 
coefficients. The transformation is valid as long as $|\Gamma t|\ll 1$, with $\Gamma$ the growth/damping rate of the waves due to the wind/dissipation. Approximate rogue wave solutions of the equation  are 
presented and discussed. The results shed some lights on the effects of 
wind and dissipation on the formation of rogue waves. 
\end{abstract}


\end{frontmatter} 



\section{Introduction} \label{sec:intro}
Modulational instability, also known as the Benjamin-Feir instability in 
the water wave community, has been discovered in the late sixties independently by Benjamin and Feir \citep{BF67} and
Zakharov \citep{zakharov68} (see \cite{zakharov2009modulation}
for an historical review on the subject and possible applications).
It describes the exponential growth of an initially  sinusoidal long wave perturbation  of  a plane wave solution of the one dimensional water wave problem. For water waves the condition of instability in infinite water depth is that 
$2\sqrt{2}a_0k_0N>1$, where $a_0$ is the amplitude of the plane wave and $k_0$ is the corresponding wave number; $N=k_0/\Delta K$ is the number of waves under the perturbation of wavenumber $\Delta K$.
The modulaitonal instability is frequently studied within the 
Nonlinear Schr\"odinger (NLS) equation that describes weakly
 nonlinear and dispersive waves in the narrow band approximation.
In this context, the nonlinear stages of the modulational instability
are described by exact solutions of the NLS, known as 
Akhmediev breathers
\citep{akhmediev1987exact,peregrine1983water}. 
Other exact NLS solutions which describe the focussing of 
an initially non-small perturbation have been derived 
in \citep{ma1979perturbed,kuznetsov1977solitons}.
Such solutions have been considered as prototypes of rogue waves 
\cite{dysthe99,osborne00}.

Within the one dimensional NLS equation, the modulational instability is well understood. What is probably less clear  is the modulation of waves and the formation of rogue waves in forced (by wind) or damped (by dissipation) conditions. In this regard in the past there has been a number of experimental works, \cite{bliven1986experimental,hara1991frequency,waseda1999experimental}, 
which did not gave a clear picture on the effect of the wind on the 
modulaitonal instability. 
A careful  discussion of the discrepancy of the results presented in the above papers can be found in 
\cite{waseda1999experimental}. According to their discussion the role
of the wind is twofold: i) the wind changes the growth rate of the instability; 
 ii) the natural  selection of the sideband frequency is altered with respect to the
 no wind conditions. 

Concerning damping effects, it has been  showed  in  
\cite{segur2005stabilizing} that 
any amount of dissipation stabilizes the modulational instability, 
questioning the role of the modulational instability in the formation 
of rogue waves, \citep{segur2008can}. 
More recently, the role of dissipation and wind in the modulational instability
 has been considered together 
within the NLS equation, \cite{kharif2010modulational}  (then confirmed  
by fully nonlinear simulations, \cite{kharif2010under}). The authors  performed a linear stability analysis and numerical simulations and  found that,  
in the presence of wind, young waves are more sensitive to modulational instability than old waves. 

The  just mentioned numerical results (except the one in \cite{kharif2010under})
are all based on the following forced and damped Nonlinear Schr\"{o}dinger equation:
\begin{equation}
i\frac{\partial A}{\partial t}-\alpha\frac{\partial^2 A}{\partial x^2}-
\beta|A|^2A=i\Gamma A \label{nlsfd}.
\end{equation}
$A$ is the wave envelope, $\alpha$ and $\beta$ are 
two coefficients that depend on the wavenumber, $k_0$, of the carrier wave.
  The right-hand side is responsible for the
forcing, $\Gamma>0$, and/or dissipation, $\Gamma<0$. The two effects are additive so that $\Gamma$ is in general the sum of forcing coefficient plus a 
damping one. 
 The wind forcing  depends on the ratio between air and water density and the dissipation on 
the water viscosity, therefore the absolute value of $\Gamma$ is 
always a small quantity. Finding analytical solutions of equation (\ref{nlsfd}) is not an obvious task. In the present paper we take advantage of the smallness
 of $\Gamma$ and, after a suitable transformation, we are able to 
 find breather solutions of the forced-damped NLS equation.
 In the following sections we first describe the transformation and then present the rogue wave analytical solutions.  

\section{Reduction of the forced/damped NLS to 
the standard NLS } \label{sec:brether_sol}
We considered the NLS equation discussed in  \cite{kharif2010modulational}
\begin{equation}
i\bigg(\frac{\partial A}{\partial t}+c_g\frac{\partial A}{\partial x}\bigg)-\frac{1}{8}\frac{\omega_0}{k_0^2}\frac{\partial^2 A}{\partial x^2}-
\frac{1}{2}\omega_0 k_0^2 |A|^2A=i\Gamma A \label{nlsfdd}
\end{equation}
with
\begin{equation}
\Gamma=\frac{1}{2 g \kappa^2}\frac{\rho_a }{ \rho_w  }\gamma \omega_0\bigg(\frac{u_*}{c}\bigg)^2-2 \nu k_0^2
\end{equation}
here $\kappa$ is the Von Karman constant and $u_*$ is the friction velocity, 
$g$ is the gravity acceleration, $\rho_a$ and $\rho_w$ are 
the air and water density, respectively; $\gamma$ is a coefficient to be determined from the solution of Rayleigh equation associated to the stability of the 
wind wave problem (see also 
\cite{leblanc2007amplification} for a justification of the wind forcing term); $c$ is the phase velocity, $\nu$ is the water kinematic viscosity. In \cite{kharif2010modulational}
the equation is written in a nondimensional form and the coefficient $K=\Gamma/\omega_0$
is introduced). 
The surface elevation is related to the envelope as follows:
\begin{equation}
\eta(x,t)=\frac{1}{2}\bigg(A(x,t)\exp[i(k_0x-\omega_0 t)]+c.c\bigg) \, .
\end{equation}
Note that we use a different definition of the surface 
elevation from the one in \cite{kharif2010modulational}
where the 1/2 factor is not included
(the consequence is that the coefficient in the nonlinear term in 
equation (\ref{nlsfdd})
differs by a factor of 4 from the one in equation (3.1) in  \cite{kharif2010modulational}). If $\epsilon$ is the small parameter in the derivation of the NLS, 
then it is assumed that the right-hand side term in 
(\ref{nlsfdd}) is of the order of $\epsilon^2$ as the nonlinear and 
the dispersive term.

We consider the following new variable: 
\begin{equation}
B(x,t)=A(x,t)e^{-\Gamma t}
\end{equation}
and by selecting a coordinate system moving with the group velocity we get:
\begin{equation}
i\frac{\partial B}{\partial t}-\alpha\frac{\partial^2 B}{\partial x^2}-
\beta \exp^{2 \Gamma t}|B|^2B=0 \label{nlsfd2}
\end{equation}
were $\alpha$ and $\beta$ are the coefficients of the dispersive and nonlinear term, respectively. 
Written in the above form the effect of the forcing/damping term 
enters as a factor in front of  nonlinear term and has the role
of enhancing/decreasing the nonlinearity of the system as the wave evolve in 
time. 
Recalling that $ \Gamma $ is usually small,
we Taylor expand the exponential and re-write the equation as follows:
\begin{equation}
i\frac{\partial B}{\partial t}-\alpha\frac{\partial^2 B}{\partial x^2}-
\beta p(t) |B|^2B=0 \label{NLS_fd}
\end{equation}
with $p(t)={1}/{(1-2 \Gamma t)}$.
Let's introduce the following change of coordinates:
\begin{equation}
\chi(x,t)=p(t) x, \;\;\;\;\;\;\;\; \tau(t)= p(t) t
\end{equation} 
and scale the wave envelope function $B$ as follows
\begin{equation}
\psi(\chi,\tau)=\frac{B(x,t)}{\sqrt{p(t)}}
\exp\left[-i\left(\frac{\Gamma p(t) x^2}{2 \alpha}\right) \right] \, .
\end{equation}
After the transformation, the equation (\ref{nlsfd2}) results in:
\begin{equation}
i\frac{\partial \psi}{\partial \tau}-\alpha\frac{\partial^2 \psi}{\partial\chi^2}-
\beta |\psi|^2 \psi=0 \label{timeNLS}
\end{equation}
i.e., the NLS equation with constant coefficients. We have transformed the forced/damped NLS equation into the standard NLS equation whose  solutions 
can be studied analytically. From a physical point of view the transformation (and consequently the validity of 
the solutions)  is valid as long as  $2|\Gamma t| \ll 1$ (the transformation is 
singular for  $2|\Gamma t| =1$). 
We underline that the transformation of the forced/damped NLS equation to the standard one has been possible only for 
 $1/p(t)$ equal to a linear function in $t$.  For other functional dependences, the transformation does not seem to be possible.
Our result is consistent with ones reported in \cite{tian2011controllable,zhang2005variable} where 
analytical solutions of the variable coefficient NLS equation are described.

\section{Rogue wave solutions} \label{sec:rogue}
In the following we will present three analytical solutions 
corresponding to the Peregrine, the Akhmediev and the Kuznetsov-Ma 
breathers for the standard NLS.\\

\begin{figure}
\begin{center}
\includegraphics[width=12cm]{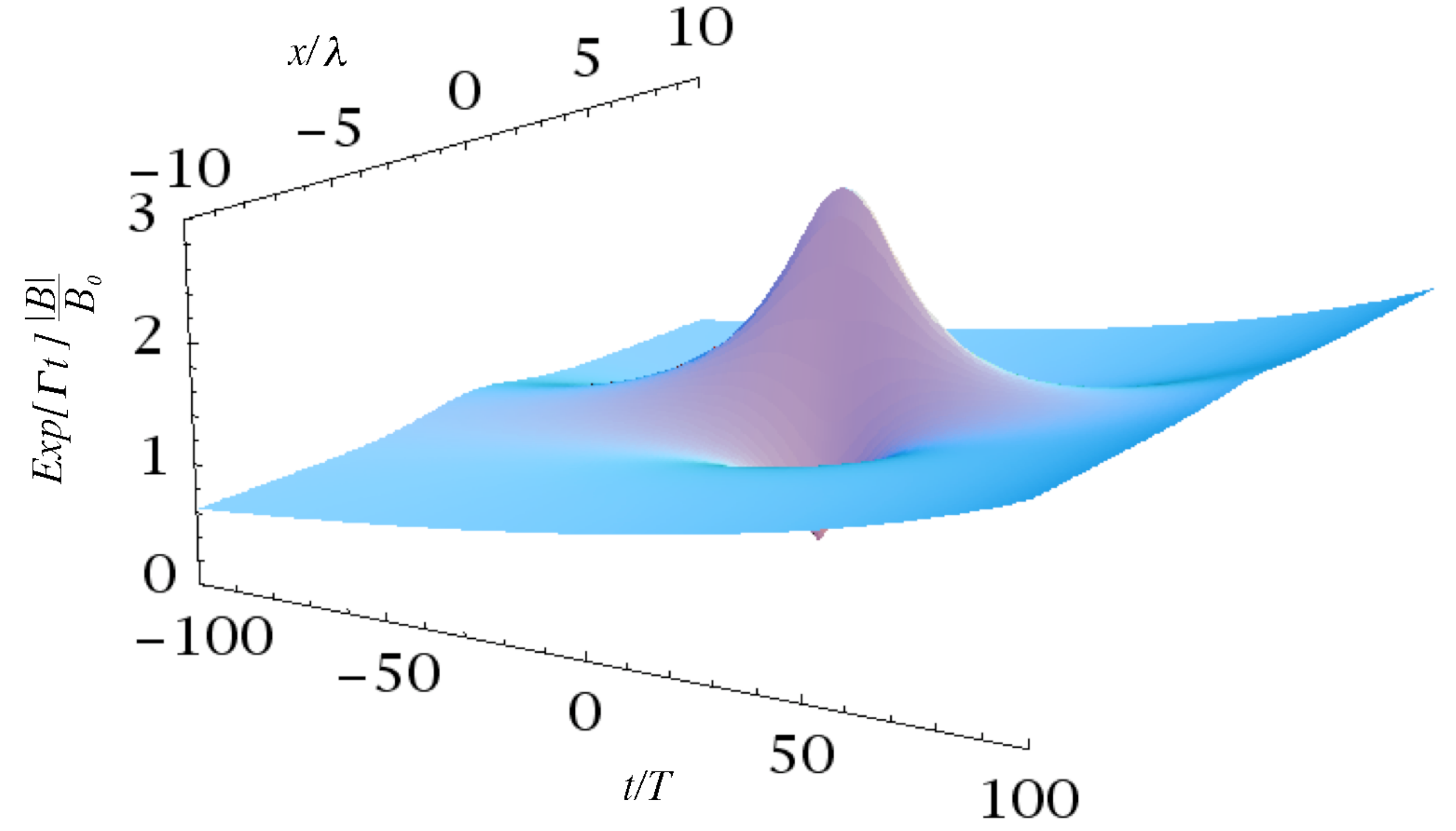}
\caption{The Peregrine solution of the forced NLS equation.
\label{fig:peregrine}}
\end{center}
\end{figure}

{\it The Peregrine  solution} also known as 
rational solution, has been originally proposed in 
\cite{peregrine1983water}. It has the peculiarity of being not periodic in time and in space: it is a wave that
``appears out of nowhere and disappears without trace'' \citep{akhmediev2009waves,shrira2010makes}; its maximum amplitude reaches three times
the amplitude of the unperturbed waves. For the above reasons it has been considered as special prototype of freak wave, \cite{shrira2010makes}.
The Peregrine solution has been recently reproduced experimentally in wave tank laboratories \citep{chabchoub2011rogue} and in optical fibers \citep{kibler2010peregrine}.
Below we present an exact analytical solution of equation 
(\ref{NLS_fd}) which is the analogous of the Peregrine solutions but for
the forced/damped case:
\begin{equation}
B(x,t)=B_0G(x,t)
\left(\frac{4(1-i 2 \beta B_0^2  p(t) t)}
{\alpha+\alpha(2 \beta B_0^2  p(t) t)^2 +2 \beta B_0^2 (p(t) x)^2}
   -1\right)   
\end{equation}
with
\begin{equation}
G(x,t)=\sqrt{p(t)}\exp\left[i\left(\frac{\Gamma p(t) x^2}{2 \alpha} -\beta B_0^2 p(t) t\right) \right] \, .
\label{G}
\end{equation}
In figure \ref{fig:peregrine} we show an example of such solution
for steepness 0.1 and forcing coefficient $K=0.0004$ (the same value has been used in \cite{kharif2010modulational}). The axis are normalized by the wave period, the wavelength and the initial wave amplitude $B_0$.
 The effect of the wind/dissipation is to increase/reduced the amplitude of the plane wave. As in the case of the standard NLS, the wave appears only once in time and space. 

\begin{figure}
\begin{center}
\includegraphics[width=12cm]{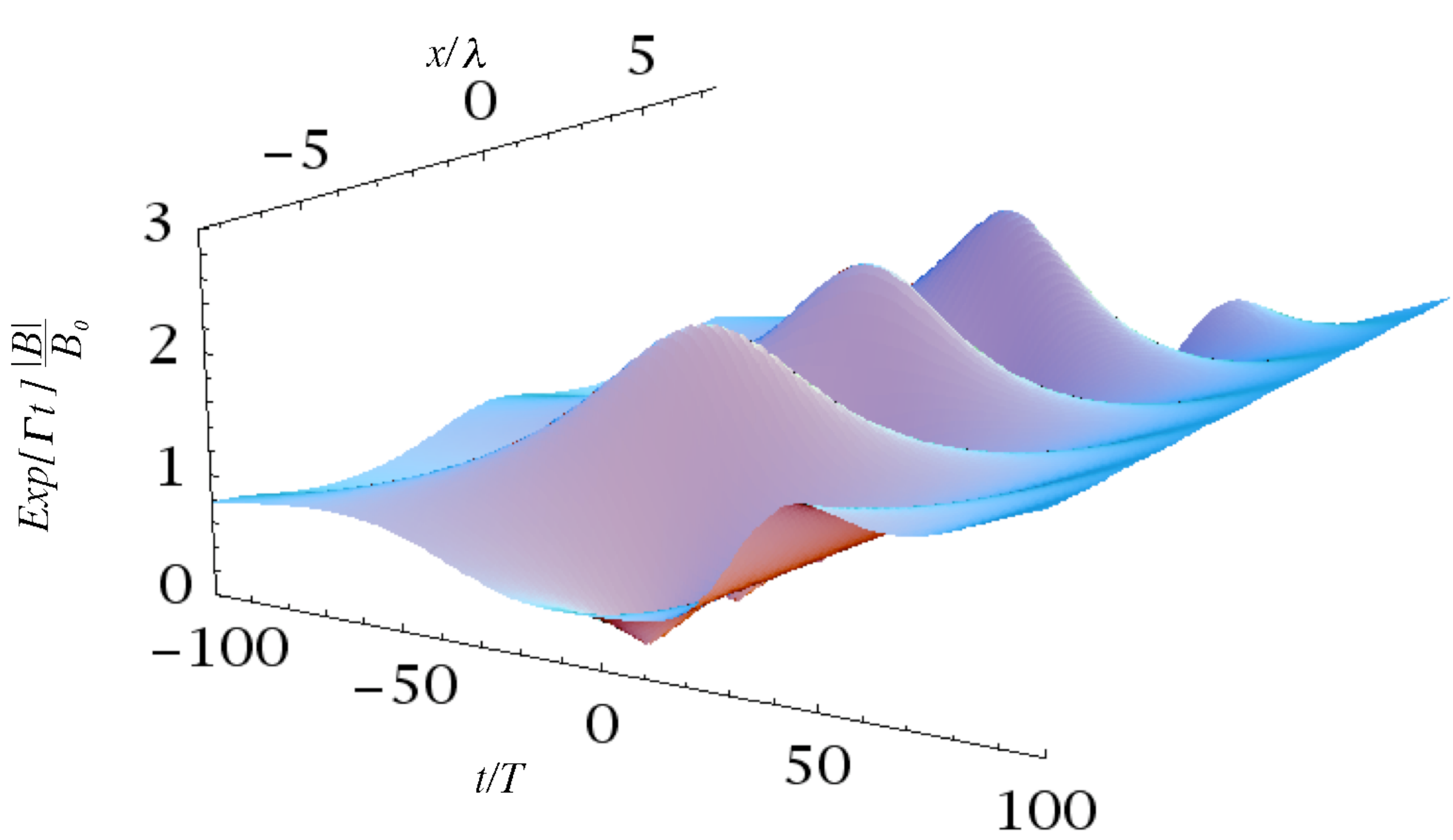}
\caption{The Akhmediev solution of the forced NLS equation.
\label{fig:akhmediev}}
\end{center}
\end{figure}

{\it The Akhmediev  solution} \cite{akhmediev1987exact} describes the modulational instability in its nonlinear regime; it is periodic in space.  It is characterized by an amplification factor which ranges from 1 to 3 (this last value corresponds to the Peregrine solution).
In the presence of a forcing or damping, the breather has the following analytical form:
\begin{equation}
B(x,t)=
B_0G(x,t)
\left(\frac{\sqrt{2}\tilde\nu^2\cosh[\sigma p(t) t]-i\sqrt{2} \tilde\sigma \sinh[\sigma p(t) t)]}
{\sqrt{2}\cosh[\sigma p(t) t]-\sqrt{2-\tilde\nu^2} \cos[\nu p(t) x]}
   -1\right)
\end{equation}
and
\begin{equation}
\nu=\frac{k_0}{N},\;\;\
\tilde\nu=\frac{\nu}{B_0}\sqrt{\frac{\alpha}{\beta}},\;\;\
\tilde\sigma=\tilde\nu\sqrt{2-\tilde\nu^2},\;\;\
\sigma=\beta B_0^2 \tilde\sigma \, .
\end{equation}
The function $G(x,t)$ is reported (\ref{G}).
It should be noted that the function is periodic in space with a
period that changes in time. 
In figure \ref{fig:akhmediev} we show an example of such solution
for steepness 0.1, $N=5$ and forcing coefficient of $K=0.0004$.  \\

\begin{figure}
\begin{center}
\includegraphics[width=12cm]{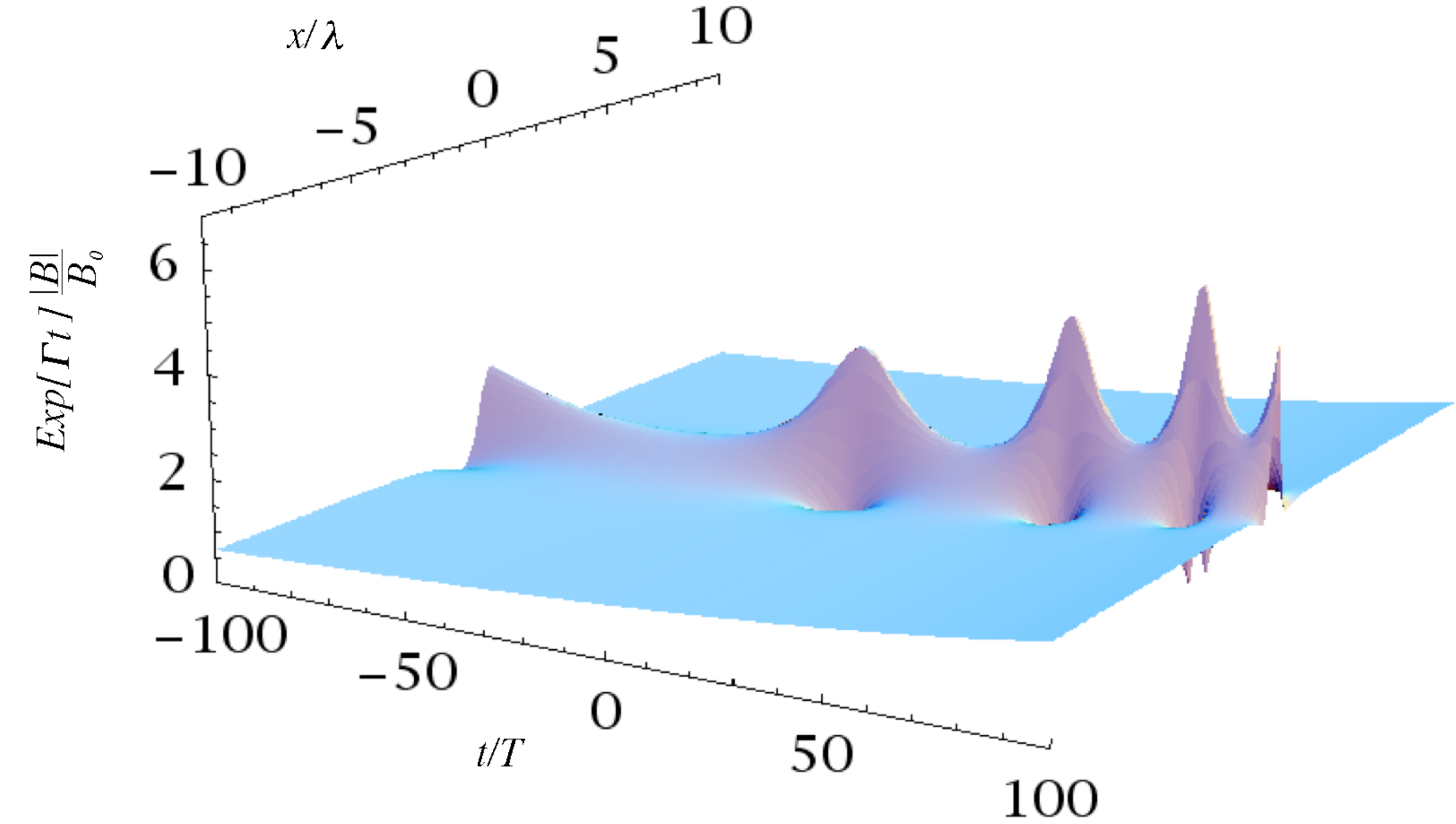}
\caption{The Kuznetsov-Ma solution of the forced NLS equation.
\label{fig:ma}}
\end{center}
\end{figure}

{\it The Kuznetsov-Ma solution} \citep{ma1979perturbed} is periodic in time  
and decrease exponentially in space. While for
the Akhmediev breather the large time (positive or negative) limit is a plane wave plus a small perturbation, 
the modulation for the Ma breather is never small.
The solution for the forced/damped equation is here reported:
\begin{equation}
B(x,t)=
B_0G(x,t)
\left(\frac{-\sqrt{2}\tilde\mu^2\cos[\rho p(t) t]+i\sqrt{2} \tilde\rho \sin[\rho p(t) t)]}
{\sqrt{2}\cos[\rho p(t) t]-\sqrt{2+\tilde\mu^2} \cosh[\nu p(t) x]}
   -1\right)
\end{equation}
with
\begin{equation}
\mu=B_0\tilde\mu\sqrt{\frac{\beta}{\alpha}},\;\;\
\tilde\rho=\tilde\mu\sqrt{2+\tilde\mu^2},\;\;\
\rho=\beta B_0^2 \tilde\rho \, .
\end{equation}
$\tilde\mu$ is a parameter related to the amplification factor. 
In figure \ref{fig:ma} we show an example of such solution
for steepness 0.1, $\tilde\mu=\sqrt{2}$ and forcing coefficient of $K=0.0004$. 
The periodicity (appearance of maxima) changes in time and increase in the presence of forcing and decrease for the damping case. 
%
%

\section{Conclusion} \label{sec:conclusion}
In the present Letter we have considered the problem of generation 
of rogue waves in the presence of wind forcing or dissipation. 
Our work is based on the one dimensional forced/damped NLS equation.

Under the assumption of $ 2|\Gamma t| \ll 1 $, where $ \Gamma $ is the forcing ($ \Gamma>0 $) 
or damping ($ \Gamma<0 $) term, we have shown how the equation can be mapped
in the standard NLS equation with constant coefficients.
In this framework, we have found explicit analytical breather solutions.


As mentioned the effect of wind/dissipation
is to increase/reduce in time the coefficient in front of the nonlinear term.
This has an impact on the modulational instability;
in particular, an initially stable (unstable) wave packet could be destabilized (stabilized) by the wind
(dissipation). Similar results have been obtained for the interaction of 
waves and current (see \cite{PhysRevLett.107.184502}).
The present results should be tested in wind waves tank facilities.

{\bf Acknowledgments}
The E.U. project EXTREME SEAS (SCP8-GA-2009-234175) is acknowledged. 
M.O. thanks Dr. GiuliNico for discussions and
ONR (grant N000141010991) for support.
We are thankful to J. Dudley for pointing us out 
reference [18].





\bibliographystyle{elsarticle-num}
\bibliography{references}







\end{document}